\documentclass{endm}
\usepackage{endmmacro,graphicx}
\usepackage{mathrsfs,hyperref}
\usepackage{amsmath,amsfonts,amssymb}

\newcommand{ \CP} { \mathcal{CP} }
\newcommand{ \ECP} { \mathcal{ECP} }

\newcommand{ \A} { \mathscr{A} }

\begin{document}

\begin{verbatim}\end{verbatim}\vspace{2.5cm}

\begin{frontmatter}

\title{Polyhedral results for the Equitable Coloring Problem}

\author[a]{I. M\'endez-Diaz\thanksref{GRANT}}
\author[b]{G. Nasini\thanksref{GRANT}}
\author[b]{D. Sever\'in\thanksref{GRANT}}

\address[a]{ FCEyN, Universidad de Buenos Aires,
	Argentina, \texttt{imendez@dc.uba.ar} }

\address[b]{ FCEIA, Universidad Nacional de Rosario,
	Argentina, \texttt{\{nasini, daniel\}@fceia.unr.edu.ar} }

\thanks[GRANT]{Partially supported by grants UBACyT X143 (2008-2010), PID-CONICET 204 (2010-2012)
and PICT 2006-1600.}

\begin{abstract}

In this work we study the polytope associated with a 0/1 integer programming formulation for the
Equitable Coloring Problem. We find several families of valid inequalities and derive sufficient
conditions in order to be facet-defining inequalities. 
We also present computational evidence of the effectiveness of including these inequalities as
cuts in a Branch \& Cut algorithm. 
\end{abstract}


\end{frontmatter}


\vspace{-2 mm}
\section{Introduction and preliminary results} \label{SINTRO}

The \emph{Equitable Coloring Problem} (ECP), originally presented in \cite{MEYER}, is a variation of the widely studied 
\emph{Graph Coloring Problem} (GCP) with additional constraints imposing that any pair
of color classes has to differ in size by at most one.
Further references and applications can be seen in \cite{KUBALE}.

A \emph{$k$-coloring} of a graph $G=(V,E)$ is a partition of $V$ in $k$ stable sets, $C_j$, with
$1 \leq j \leq k$. The stable set $C_j$ is the \emph{class of color} $j$.
An \emph{equitable $k$-coloring} (or just $k$-eqcol) of $G$ is a $k$-coloring satisfying the
\emph{equity constraints},  i.e. 
$\lfloor n / k \rfloor \leq |C_j| \leq \lceil n / k \rceil$
for each $1\leq j \leq k$, where $n = |V|$.

Unlike GCP, a graph admiting a $k$-eqcol may not admit a $(k+1)$-eqcol. 
This leads us to define $\A(G)$ as the set of $k \leq n$ such that $G$ does not admit any $k$-eqcol.
For instance, $\A(K_{3,3}) = \{1,3\}$.

The \emph{equitable chromatic number} of $G$, $\chi_{eq}(G)$, is the
minimum $k$ for which $G$ has a $k$-eqcol. Computing 
$\chi_{eq}(G)$ for arbitrary graphs
is an $NP$-hard problem \cite{KUBALE}. 

Although many integer programming formulations are known for GCP, as far as we know, just
two of these models were adapted for ECP. One case is the model in \cite{REPRESENTATIVES},
adapted in \cite{BYCBRA}.
Preliminary results concerning a Branch \& Cut algorithm based on one of the models in \cite{BCCOL} were presented in \cite{INFORMS}.  The algorithm 
turns out to be 
competitive compared to the one presented in \cite{BYCBRA}. 
This encouraged us to delve into a polyhedral
study with the aim of finding strong inequalities that allow us to improve the performance of our algorithm.


\vspace{-3 mm}
\section{The polytope $\ECP$} \label{SPOLYT}
\vspace{-1 mm}

From now on, we assume that $G$ is a graph with $n$ vertices such that $n \geq 5$ and $2 \leq \chi_{eq}(G) \leq n - 2$. Other cases are trivial. 

In \cite{BCCOL}, colorings of $G$ are identified with binary vectors $(x,w) \in \{0, 1\}^{n^2 + n}$
where $x \in \{0, 1\}^{n^2}$ and $w \in \{0, 1\}^n$, satisfying the following constraints:

\vspace{-6 mm}
{\footnotesize{
\begin{align} 
  & \sum_{j = 1}^n x_{vj} = 1 & &\forall~ v \in V                   &  \mbox {(assign a unique color to each vertex)}\nonumber \\
  & x_{uj} + x_{vj} \leq w_j  & &\forall~ uv \in E,~ j = 1,\ldots,n &  \mbox {(adjacent vertices do not share the same color)}\nonumber \\
  & w_{j + 1} \leq w_j        & &\forall~ j = 1,\ldots,n-1. &  \mbox {(eliminate some symmetric colorings)} \nonumber 
\end{align} 
}}
\vspace{-6 mm}

\noindent where $x_{vj} =1$ if color $j$ is assigned to vertex $v$ and $w_j = 1$ if color $j$ is used, i.e. $C_j \neq \varnothing$.
The \emph{coloring polytope} $\CP$ is defined as the convex hull of 
colorings of $G$. In this work, equitable colorings are identified with binary vectors defining colorings which also satisfy

\vspace{-5 mm}
{\footnotesize{
\begin{align}
  & x_{vj} \leq w_j, & \forall~ v~ \textrm{isolated},~ j = 1, \ldots, n, \label{RISOL} \\
  & \sum_{k=j}^n
         \biggl \lfloor \frac{n}{k} \biggr \rfloor \bigl( w_k - w_{k+1} \bigr) \leq \sum_{v \in V} x_{vj} \leq \sum_{k=j}^n
         \biggl \lceil \frac{n}{k} \biggr \rceil \bigl( w_k - w_{k+1} \bigr),
	 & \forall~ j = 1,\ldots, n - 1, \label{RLOWERUPPER} 
  \end{align}
}}
\vspace{-5 mm}

\noindent where $w_{n+1}$ is a dummy variable set to 0, constraints (\ref{RISOL}) ensure that isolated vertices use enabled colors
and (\ref{RLOWERUPPER}) are the equity constraints.  
The \emph{Equitable Coloring Polytope} $\ECP$ is the convex hull of the equitable colorings of $G$.

Next we state the main results related to the polyhedral structure of $\ECP$.
\vspace{-5 mm}
\begin{proposition}
The dimension of $\ECP$ is $n^2 - (|\A(G)| + 2)$.
\end{proposition}

In \cite{BCCOL}, \emph{clique inequalities} and \emph{block inequalities} are proven to be facet-defining
inequalities of $\CP$. In our case, we have:

\begin{proposition}
(i)~Let $j \leq n - 1$ and $Q$ be maximal clique of $G$ such that $|Q| \geq 2$. Then, the clique inequality
$\sum_{v \in Q} x_{vj} \leq w_j$ defines a facet of $\ECP$.\\
(ii)~Let $v \in V$ and $j \leq n - 2$. Then, the block inequality $\sum_{k=j}^n x_{vj} \leq w_j$
is valid for $\ECP$ and defines a facet of $\ECP$ if $j-1 \notin \A(G)$.
\end{proposition}

By lifting \emph{rank inequalities} and \emph{neighborhood inequalities}, also studied in \cite{BCCOL},
we obtain new families of valid inequalities which often define facets.

\begin{proposition} \label{T2RANK}
Let $j \leq n - 1$,  $S \subset V$ with $\alpha(S) = 2$ and
$Q=\{q: q\in S ,\;  S\subset N[q]\}$. Then, the \emph{$(S,Q)$-2-rank inequality} defined as
\vspace{-1 mm}
{\small \begin{equation*}
\sum_{v \in S\backslash Q} x_{vj} + 2 \sum_{v \in Q} x_{vj} \leq 2 w_j.
\end{equation*} \vspace{-4 mm}}

\noindent is valid for $\ECP$.
Let us assume that $|Q|\geq 2$ and no connected component of the complement graph of $G[S \backslash Q]$
is bipartite. The inequality defines a facet of $\ECP$ if one of the following conditions holds:
\begin{itemize}
\item for all $v \in V \backslash S$,  $Q \cup \{v\}$ is not a clique,
\item $n$ is odd, $j \leq \lceil n/2 \rceil - 1$ and for all
$v \in V \backslash S$ such that $Q \subset N(v)$, there exists a stable set $H$ of size 3 such that
$v\in H$ and $|H\cap S| =2$, and the complement of $G - H$ has a perfect matching,
\item $n$ is even, $j \leq \lceil n/2 \rceil - 1$ and for all
$v \in V \backslash S$ such that $Q \subset N(v)$, 
there exist two disjoint
stable sets of size 3, $H$ and $H'$, such that $v\in H$ and $|H\cap S| =2$, 
and the complement of $G - (H \cup H')$ has a perfect matching.
\end{itemize}
\end{proposition}
If $Q=\varnothing$ or $Q = \{q\}$, the $(S,Q)$-2-rank inequality is respectively dominated by
the inequalities
\vspace{-1 mm}
{\small \begin{equation*}
\sum_{v \in S} x_{vj} + \sum_{v \in V} x_{vn-1} \leq 2 w_j + w_{n-1} - w_n,~~\textrm{or}
\end{equation*} \vspace{-3 mm}
\begin{equation*}
\sum_{v \in S\backslash\{q\}} x_{vj} + 2 x_{qj} + x_{qn} \leq 2 w_j
\end{equation*} \vspace{-3 mm}}

\noindent
which also usually define facets of $\ECP$.

\begin{proposition} \label{TNEIGHBOR1}
Given $j \leq n - 1$, $u \in V$ and $S \subset N(u)$ with $\alpha(S)\geq 2$, the \emph{$(u,j,S)$-subneighborhood inequality} defined as

\vspace{-3 mm}
{\small \begin{equation*} 
\gamma_{jS} x_{uj} + \sum_{v \in S} x_{vj} +
\sum_{k = j+1}^n (\gamma_{jS} - \gamma_{kS}) x_{uk} \leq \gamma_{jS} w_j,
\end{equation*} \vspace{-3 mm}}

\noindent  
where $\gamma_{kS} = \min \{\lceil n/k \rceil, \alpha(S)\}$, is a valid inequality for
$\ECP$. If $S = N(u)$ or $\alpha(S) \leq \lceil n/j \rceil - 1$, the inequality defines a facet of $\ECP$ when the following conditions hold:
\begin{itemize}
\item 
for all $k \in \{ \lceil \frac{n}{i} \rceil - 1 : 2 \leq i \leq \gamma_{jS} - 1 \}$, 
there exists a $k$-eqcol such that $|C_j\cap S|= \gamma_{kS}$,
\item for all $v \in N(u) \backslash S$, there exists an equitable coloring such that
$|C_j \cap S| = \alpha(S)$ and $(C_j \cap N(u)) \backslash S=\{v\}$.
\end{itemize}
\end{proposition}

Finally, we obtain three new families of valid inequalities for $\ECP$, which were
not derived from any of the valid inequalities given in \cite{BCCOL}.

\begin{proposition} \label{TONLYCOLORS}
Let $S \subset \{1,\ldots,n\}$. The \emph{$S$-color inequality} defined as

\vspace{-3 mm}
{ \small \begin{equation*} 
\sum_{j \in S} \sum_{v \in V} x_{vj} \leq \sum_{k=1}^{n} b_{Sk} (w_k - w_{k+1}),
\end{equation*} \vspace{-3 mm}}

\noindent where $d_{Sk} = |S \cap \{1,\ldots,k\}|$ and $b_{Sk} = d_{Sk} \lfloor \frac{n}{k} \rfloor + \min \{ d_{Sk}, n - k \lfloor \frac{n}{k}\rfloor \}$, 
is a valid inequality for $\ECP$.
In addition, if $3 \leq |S| \leq n - 2$, $S$ contains all the colors greater than 
$n - \lceil \frac{|S|+1}{2} \rceil$ and the complement of $G$ has a matching
of size $\lceil \frac{|S|+1}{2} \rceil$, then the $S$-color inequality defines a facet of $\ECP$.
\end{proposition}

\begin{proposition} \label{TNEIGHBOR2}
Given $u$ a non universal vertex of $G$ and
$j \leq \lfloor n/2 \rfloor$ such that $\alpha(N(u)) \geq \lfloor n/j \rfloor$,
the \emph{$(u,j)$-outside-neighborhood inequality} defined as

\vspace{-7 mm}
{\small \begin{equation*} 
(\lfloor n/j \rfloor - 1) x_{uj} - \sum_{v \in V \backslash N[u]} x_{vj}
+ \sum_{k = j+1}^n b_{jk} x_{uk} \leq \sum_{k = j+1}^{n} b_{jk} (w_k - w_{k+1}),
\end{equation*} \vspace{-5 mm}}

\noindent where $b_{jk} = \lfloor n/j \rfloor - \lfloor n/k \rfloor$, is valid for $\ECP$ and defines
a facet of $\ECP$ if the following conditions hold:
\begin{itemize}
\item
there exists $v\in V \backslash N[u]$ such that $N(u) \backslash N(v)\neq \varnothing$, 
\item 
if $n$ is odd, the complement of $G - u$ has a perfect matching,
\item for all $v \in V \backslash N[u]$, there exists a $\lfloor n/2 \rfloor$-eqcol such that $C_j = \{u, v\}$,
\item for all $k$ such that $j \leq k \leq \lfloor n/2 \rfloor$ and $\lfloor \frac{n}{k} \rfloor > \lfloor \frac{n}{k+1} \rfloor$,
      there exists a $k$-eqcol such that $|C_j \cap N(u)| = \lceil n/k \rceil$, and a $k$-eqcol such that
      $u \in C_j$ and $|C_j \backslash N[u]| = \lfloor n/k \rfloor - 1$,
\item for all $k \in \{j, \ldots, n - 3\} \backslash \A(G)$, there exists a $k$-eqcol lying on the face defined by the inequality.
\end{itemize}
\end{proposition}

\begin{proposition} \label{TNEIGHBOR3}
Given $u \in V$, $Q$ be a clique of $G$ such that
$Q \cap N[u]=\varnothing$ and $j,k$ such that $j \leq k \leq n - 2$ and 
$\alpha(N(u)) \geq \lceil n/k \rceil - 1$. The \emph{$(u,j,k,Q)$-clique-neighborhood inequality} defined as

\vspace{-7 mm}
{\small \begin{multline*} 
\bigl(\lceil n / k \rceil - 1 \bigr) x_{uj} + \sum_{v \in N(u)\cup Q} x_{vj} + 
 \sum_{l = k+1}^n \bigl( \lceil n / k \rceil - \lceil n / l \rceil \bigr) x_{ul} + \sum_{v \in V} x_{vn-1}
+ \sum_{v \in V \backslash \{u\}} x_{vn} \\
\leq \sum_{l = j}^{k-1} b_{ul} (w_l - w_{l+1})
+ \sum_{l = k}^{n-2} \lceil n/k \rceil (w_l - w_{l+1})
+ \sum_{l = n-1}^{n} (\lceil n/k \rceil + 1) (w_l - w_{l+1}),
\end{multline*} \vspace{-4 mm}}

\noindent where $b_{ul}=\min\{\lceil n / l \rceil, \alpha(N(u)) + 1\}$, is a valid inequality for $\ECP$.
If there exists $v \in Q$ such that $N(u) \backslash N(v)\neq \varnothing$, the inequality defines a
facet of $\ECP$ when the following conditions hold:
\begin{itemize}
\item for all $l \in \{j, \ldots, n-3\} \backslash \A(G)$, there exists an
$l$-eqcol lying on the face defined by the inequality,
\item for all $v \in V \backslash (N[u] \cup Q)$, there exist two $k$-eqcols lying on the face
defined by the inequality, with  $v \in C_j$ in the first one and where the second one is
obtained from the first by only changing the color of $v$,  i.e.  $v \notin C_j$,
\item for all $1 \leq i \leq \lceil n/k \rceil - 1$, if
$l = \max \{\lceil \frac{n}{i} \rceil - 1, n-2 \}$, there exist two $l$-eqcols lying on the face
defined by the inequality such that $u \in C_j$ in one of them and $u \in C_l$ in the other.
\end{itemize}
\end{proposition}

Although  the sufficient conditions in the previous results are strong, we find several cases where
they hold. Moreover, even when the inequalities do not define facets, the dimension of the faces
defined by them is quite high. For example, if $k \leq \lceil n/2 \rceil - 1$, it can be proved
that the dimension of the face defined by the $(u,j,k,Q)$-clique-neighborhood inequality is at least
$dim(\ECP) - \bigl(3n - |\A(G)| - \lfloor n/2 \rfloor - |N(u)| - |Q| - 5 \bigr)$.


\vspace{-2 mm}
\section{Computational performance of valid inequalities} \label{SBYCALG}
\vspace{-1 mm}

In this section, we report on the computational performance of the families of valid inequalities
studied in the previous section, embedded as cuts in a B\&C algorithm for solving ECP.

In order to strengthen the formulation and avoid considering classes of symmetric colorings, constraints
$x_{vj} = 0,~ \forall~ 1 \leq v < j \leq n$ are considered within the initial relaxation, and
$ x_{vj} \leq \sum_{u = j-1}^{v-1} x_{u j-1},~ \forall~ 2 \leq j \leq v \leq n$ are handled as cuts
during the optimization.

The cutting process consists in looking for violated clique and $(S,Q)$-2-rank inequalities with a greedy algorithm.
During the separation of clique inequalities, it attempts to find violated
$(u,j,k,Q)$-clique-neighborhood inequalities by scanning vertices $u$ not adjacent to a given clique
$Q$. Whenever not enough cuts were generated, it tries to add block,
$(u,j,N(u))$-subneighborhood and $(u,j)$-outside-neighborhood inequalities, handled by
enumeration, and $S$-color inequalities with a greedy algorithm.
Separation routines for clique and block inequalities are exposed in \cite{BCCOL}.
The B\&C algorithm also includes an initial heuristic, a primal heuristic and a custom branching rule.

Experiments were carried out over random instances of 70 vertices with different density percentages
and 2 hours time limit. We compare our B\&C algorithm with (BC$^+$) and without (BC) our new inequalities against the general purpose IP-solver CPLEX 12.1 and
results reported in \cite{BYCBRA}.

\begin{center} { \footnotesize \renewcommand{\arraystretch}{1}\addtolength{\tabcolsep}{5pt}
\begin{tabular}{c@{\hspace{3pt}}|@{\hspace{3pt}}c@{\hspace{3pt}}c@{\hspace{3pt}}c@{\hspace{3pt}}c@{\hspace{3pt}}|@{\hspace{3pt}}c@{\hspace{3pt}}c@{\hspace{3pt}}c@{\hspace{3pt}}c@{\hspace{3pt}}|@{\hspace{3pt}}c@{\hspace{3pt}}c@{\hspace{3pt}}c@{\hspace{3pt}}c}
\%  & \multicolumn{4}{c}{\% solved inst.} & \multicolumn{4}{c}{Nodes (average)}
	& \multicolumn{4}{c}{Time in sec. (average)} \\
dens.  & BC$^+$ & BC & CPX & \cite{BYCBRA}
  & BC$^+$ & BC & CPX & \cite{BYCBRA}
  & BC$^+$ & BC & CPX & \cite{BYCBRA} \\
 
\hline
10 & 100 & 100 & 100 & 100 & 3.4 & 4 & 13.3 & 57 & 0.3 & 0.3 & 4 & 109 \\
30 & 90  & 90  & 0   & 0   & 2135 & 3949 & $-$ & $-$ & 276 & 224 & $-$ & $-$ \\
50 & 70  & 70  & 0   & 0   & 7932 & 21595 & $-$ & $-$ & 1354 & 2145 & $-$ & $-$ \\
70 & 80  & 80  & 10  & 100 & 525  & 2970 & 214 & 678 & 128 & 446 & 4380 & 273 \\
90 & 100 & 100 & 100 & 100 & 5.1 & 14.5 & 30 & 9.4 & 2.6 & 2.8 & 29 & 11 \\ 
\hline
\end{tabular} }
\end{center}

As one may appreciate from the table, the addition of our cutting planes has shown to be particularly
useful in substantially decreasing the number of Branch-and-Bound nodes and the CPU time was significantly
reduced on medium and high density instances. 


\end{document}